\documentclass[english,aps,a4paper,prd,groupedaddress,nofootinbib,preprint,floatfix]{article}

\RequirePackage[dvipsnames]{xcolor}

\usepackage{colortbl}

\usepackage{amsmath,amssymb}

\usepackage{cite}
\usepackage[a4paper, total={6in, 8in}]{geometry}

\usepackage{braket}
\usepackage{cancel}
\newcommand{\MET}{$\cancel{E}_T$}

\usepackage[utf8]{inputenc}
\usepackage{amsmath}
\usepackage{amssymb}
\usepackage{graphicx}
\usepackage{caption}
\usepackage{subcaption}
\usepackage{float}
\usepackage{multirow}
\usepackage{url}
\usepackage{hyperref}
\usepackage{cleveref}
\usepackage{indentfirst}

\usepackage{algorithm}
\usepackage{algpseudocode}

\def\lsim{\raise0.3ex\hbox{$\;<$\kern-0.75em\raise-1.1ex\hbox{$\sim\;$}}}

\hypersetup{
    colorlinks,
    linkcolor={red!50!black},
    citecolor={blue!50!black},
    urlcolor={blue!80!black}
}

\usepackage[bitstream-charter]{mathdesign}
\DeclareSymbolFont{usualmathcal}{OMS}{cmsy}{m}{n}
\DeclareSymbolFontAlphabet{\mathcal}{usualmathcal}

\makeatletter
\renewcommand\@dotsep{240}   
\makeatother

\begin{document}

\newcommand{\AddrLIP}{%
LIP -- Laborat\'orio de Instrumenta\c{c}\~ao e F\'isica Experimental de Part\'iculas, Escola de Ciências, Campus de Gualtar, Universidade do Minho, 
 4701-057 Braga, Portugal}

\newcommand{\AddrDF}{%
Departamento de F\'isica, Escola de Ciências, Campus de Gualtar, Universidade do Minho, 
 4701-057 Braga, Portugal}

\newcommand{\AddrLIPL}{%
LIP -- Laboratório de Instrumentação e Física Experimental de Partículas, Av. Prof. Gama Pinto, 2, 1649-003 Lisboa, Portugal}

\newcommand{\AddrSoton}{%
Department of Physics and Astronomy,
University of Southampton,
SO17 1BJ Southampton, United Kingdom
 }

\begin{center}{\Large \textbf{
Fitting a Collider in a Quantum Computer: Tackling the Challenges of Quantum Machine Learning for Big Datasets\\
}}\end{center}

\begin{center}
Miguel Ca\c{c}ador Peixoto\textsuperscript{1},
Nuno Filipe Castro\textsuperscript{1, 2},
Miguel Crispim Rom\~ao\textsuperscript{1, 3},\\
Maria Gabriela Jord\~ao Oliveira\textsuperscript{1} and Inês Ochoa\textsuperscript{4}

\end{center}

\begin{center}
{\bf 1} \AddrLIP
\\
{\bf 2} \AddrDF
\\
{\bf 3} \AddrSoton
\\
{\bf 4} \AddrLIPL
\\
\end{center}

\begin{center}
\today
\end{center}


\section*{Abstract}
{\bf
Current quantum systems have significant limitations affecting the processing of large datasets with high dimensionality, typical of high energy physics. In the present paper, feature and data prototype selection techniques were studied to tackle this challenge. A grid search was performed and quantum machine learning models were trained and benchmarked against classical shallow machine learning methods, trained both in the reduced and the complete datasets. The performance of the quantum algorithms was found to be comparable to the classical ones, even when using large datasets. Sequential Backward Selection and Principal Component Analysis techniques were used for feature's selection and while the former can produce the better quantum machine learning models in specific cases, it is more unstable. Additionally, we show that such variability in the results is caused by the use of discrete variables, highlighting the suitability
of Principal Component analysis transformed data for quantum machine learning applications in the high energy physics context.
}


\section{Introduction}

The Standard Model of Particle Physics (SM) provides a remarkable description of the fundamental constituents of matter and their interactions, being in excellent agreement with the collider data accumulated so far. Nonetheless,  there are still important open questions, unaddressed by the SM, such as gravity, dark matter, dark energy, or the matter-antimatter asymmetry in the universe~\cite{Ellis:2012zz}, motivating a comprehensive search program for new physics phenomena beyond the SM (BSM) at the Large Hadron Collider (LHC) at CERN.


The search for BSM phenomena at colliders poses specific challenges in data processing and analysis, given the extremely large datasets involved and the low signal to background ratios expected. In this context, the analysis of the collision data obtained by the LHC experiments often relies on machine learning (ML), a field in computer science that can harness large amounts of data to train generalizable algorithms for a variety of applications \cite{guest2018deep, Feickert:2021ajf}, such as classification tasks. These techniques have shown an outstanding ability to find correlations in high-dimensional parameter spaces to discriminate between potential signal and background processes. They are known to scale with data, and usually rely on a large number of learnable parameters to achieve their remarkable performance.

In order to train these large models, classical\footnote{\emph{Classical} is used throughout the paper as opposed to \emph{quantum} machine learning.} machine learning (CML) takes advantage of hardware accelerators, such as graphics processing units (GPUs), 
for efficient, parallel, and fast matrix multiplications. On the other hand, a new class of hardware is becoming available, with the advent of noisy intermediate-scale quantum (NISQ) computing devices. This accelerated the development of new quantum algorithms targeted at exploiting the capacity and feasibility of this new technology for ML applications.

Quantum machine learning (QML) is an emerging research field aiming to use quantum circuits to tackle ML tasks. One of the motivations for using this new technology in high energy physics (HEP) relates to the intrinsic properties of quantum computations, namely representing the data in a Hilbert space  where the data can be in a superposition of states or in entangled states, which can allow to explore additional information in data analysis and, eventually, contribute to better classification of HEP events, namely in the context of the search for BSM phenomena. Recently, this new technology has been applied to various HEP problems\cite{Guan_2021}. Namely, in event reconstruction~\cite{shapoval2019quantum, tuysuz2020particle, bapst2020pattern, funcke2022studying, zlokapa2021charged, das2019track, wei2020quantum}, classification tasks ~\cite{mott2017solving, zlokapa2021quantum, terashi2021event, wu2021application,  blance2021quantum,
 gianelle2022quantum, chen2022quantum, belis2021higgs, Araz_2022} data generation ~\cite{rehm2023quantum, chang2021quantum, refId0, delgado.106.096006, Borras_2023} and anomaly detection problems \cite{Alvi_2023, woźniak2023quantum, ngairangbam2022anomaly,schuhmacher2023unravelling}. 

Despite the promising potential of quantum computation, NISQ processors have important limitations, such as the qubit quality (\emph{i.e.} the accuracy with which it is possible to execute quantum gates), the qubit lifetime and the limited depth of quantum circuits, since for large circuits the noise overwhelms the signal~\cite{Preskill_2018,qubitmapping}. This necessarily limits the complexity of the circuits and the size of the datasets used to train them.

In this paper we perform a systematic comparison of the performance of QML and shallow CML algorithms in HEP. The choice to focus on shallow methods rather than state-of-the-art architectures based on deep neural networks is to provide a fair comparison between methodologies, since neural networks are known to require large datasets (both in terms of sample size and dimension) to achieve good performance, something that is not feasible with current quantum computers. By choosing CML algorithms suited for smaller datasets, we will add to the on-going discussion regarding potential advantages of quantum computing by comparing QML and CML in the same footing.

The use of QML algorithms in this context is studied by targeting a common binary classification task in HEP: classifying a BSM signal against SM background. A benchmark BSM signal leading to the $Z t$ final state is considered, in events with multiple leptons and $b$-tagged jets, which can be used to achieve a reasonable signal to background ratio. Variational quantum classifiers (VQC) are trained and optimized via a grid search. The use of reduced data is explored, considering both the number of features and the number of events, via different strategies: ranking of features, data transformations aiming for a richer reduced set of features, use of random samples, and choice of representative data samples.

\section{Quantum Machine Learning}

The QML algorithms are implemented using a quantum circuit, \emph{i.e.}  a collection of quantum gates applied to an $n$-qubit quantum state, followed by a measurement (or multiple measurements) that represent the output of the circuit. In order to implement a learning algorithm, the quantum circuit can be parameterized with parameters that can be learned by confronting the measurement to a loss function.

QML is effectively an extension of CML techniques to the Hilbert space, where instead of representing data as vectors in a high-dimensional real space, we encode it in state vectors of a Hilbert space. A QML algorithm, such as a quantum neural network, can be implemented using the quantum equivalent of a perceptron, one of the building blocks of CML. A problem arises from the realization that the activation functions used in CML can not be expressed using a linear operation, which is inherently required from the quantum evolution of a state. Ideas have been proposed to imitate an activation function in the quantum space, \cite{gupta2001quantum, schuld2015simulating} but, in the current paper, only variational quantum classifiers~\cite{1802.06002, Schuld_2020} are used for binary classification.

A VQC is a parameterized quantum circuit, a circuit type containing adjustable gates with tunable parameters. These gates are a universal set of quantum gates and, in the current study, rotation [$R_X(w)$, $R_Y(w)$, $R_Z(w)$] and CNOT gates are used\footnote{Even if, in general, the phase shift gate $P(w)$ should be included, this gate does not change the final outcome (\emph{i.e.} it does not impact probabilities), so it can be discarded.}.

The considered VQC pipeline used has the following components:
\begin{itemize}
    \item \textbf{Data Embedding}: the numerical vector $X$ representing the classical information is converted to the quantum space with the preparation of an initial quantum state, $\ket{\psi_X}$, which represents a HEP event.
    \item \textbf{Model circuit}: a unitary transformation $U(w)$, parameterized by a set of free parameters $w$, is applied to the initial quantum state $\ket{\psi_X}$. This produces the final state $\ket{\psi_X'} = U(w)\ket{\psi_X}$.
    \item \textbf{Measurement}: a measurement of an observable is performed in one of the qubits of the state $\ket{\psi_X'}$, which will give the prediction of the model for the task at hand. The training of a VQC aims to find the best set of parameters $w$ to match the event labels to the prediction.
\end{itemize}


Throughout this work, the \textit{PennyLane} package~\cite{bergholm2018pennylane} was used as a basis for the hybrid quantum-classical machine learning applications. Leveraging \textit{PennyLane}'s \texttt{default.qubit} quantum simulator, a straightforward tool for quantum circuit simulations, we trained and assessed the performance of various QML algorithms. Subsequently, the performance of the algorithms trained on IBM's quantum computers was gauged by integrating \textit{PennyLane} with IBM's quantum computing framework, \textit{Qiskit}~\cite{Qiskit}.

\subsection{Data Embedding}

Before passing the data through the VQC, the preparation of the initial quantum state $\ket{\psi_X}$ is required. This is called data embedding, and there are a number of proposals to perform this step \cite{larose2020robust}. Among the different possible embeddings, it was chosen to test amplitude embedding against angle embedding. The preliminary results have shown that angle embedding leads to a better performance than the former, as previously reported in a different context~\cite{gianelle2022quantum}. In this paper angle embedding was, therefore, the adopted choice. Further studies on possible embeddings is left for future works.


For an $N$-dimensional vector of classical information, $X=(x_1, x_2, ..., x_N)$, the state entering the VQC will be defined via a state preparation circuit applied to the initial state of $\ket{0}^{\bigotimes N}$. The information contained in $X$ is embedded as angles: these are the values used in rotation gates applied to each qubit, thus requiring $N$ qubits for embedding $N$ features from the original dataset.

In the current study, the embedding is done using rotations around the $x$-axis on the Bloch sphere, thus defining the quantum state embedded with the classical information as:
\begin{equation}
    \ket{\psi_X} =  \bigotimes_{i=1}^{N} R_X(x_i) =  \bigotimes_{i=1}^{N} 
    \left[
    \cos\left(\frac{x_i}{2}\right) \ket{0} - i\sin\left(\frac{x_i}{2}\right) \ket{1} 
    \right]
    \,\, ,
    \label{angle_embedding}
\end{equation}
where $R_X(x) = e^{-ix\hat{\sigma}_x}$ and $\hat{\sigma}_x$ is a Pauli operator. In this embedding each of the considered features of the original dataset is required to be bound between $[-\pi, \pi]$. 

\subsection{Model Circuit}

The model circuit is the key component of the VQC and includes the learnable set of parameters. It is defined by a parameterized unitary circuit $U(w)$, with $w$ being the set of tunable parameters, which will evolve a quantum state embedded with classical information $\psi_X$ into the final state $\psi_X'$. 

Analogously to the architecture of a classical neural network, the model circuit is formed by layers. Each layer is composed of an assemblage of rotation gates applied to each qubit in the system, followed by a set of CNOT gates. 

A rotation gate, $R$, is designed to be applied to one single qubit and rotate its state. It is composed by 3 learnable parameters: $\phi, \theta, \omega$, which enables the gate to rotate any arbitrary state to any location on the Bloch sphere.

\begin{equation}
    R(\phi,\theta,\omega) = RZ(\omega)\, RY(\theta) \, RZ(\phi)= \begin{bmatrix}
        e^{-i(\phi+\omega)/2}\cos(\theta/2) & -e^{i(\phi-\omega)/2}\sin(\theta/2) \\
        e^{-i(\phi-\omega)/2}\sin(\theta/2) & e^{i(\phi+\omega)/2}\cos(\theta/2)
        \end{bmatrix}
\end{equation}

Since all the learnable parameters of the VQC are contained inside the rotation gates, and each gate has 3 parameters, the shape of the weight vector is $w\in \mathbb{R}^{n\times l\times 3}$, where $n$ is the number of qubits of the current system and $l$ is the number of layers in the network. As mentioned in the previous section, $n$ will depend on the number of features in the data and $l$ is a hyper-parameter (HP) to be tuned.

After rotating the qubits' state, a collection of CNOT gates will be applied to entangle the qubits. The CNOT gate is a 2-qubit gate with no learnable parameters. It will flip the state of the so-called target-qubit, based on the value of the control-qubit, and it is usually represented by having two inputs as such: CNOT(control-qubit, target-qubit). Given the number of qubits, the CNOT arrangement is implemented as detailed in \autoref{algo_CNOT}.

\vspace{5pt}
\begin{algorithm}
\label{cnot_alg}
\caption{CNOT Arrangement}
\begin{algorithmic}
\Require $n \geq 2$, $n$ being the number of qubits.

\If{$n == 2$}
    \State \texttt{$CNOT(1, 0)$}
\Else
    \For{\texttt{$qubit \gets 0$ to $n-1$ }}
        \If{\texttt{$qubit == n-1$ }}
            \State \texttt{$CNOT(qubit, 0)$}
        \Else
            \State \texttt{$CNOT(qubit, qubit+1)$}
        \EndIf
    \EndFor
\EndIf
\end{algorithmic}
\label{algo_CNOT}
\end{algorithm}
\vspace{5pt}

\subsection{Measurement}

\begin{figure*}[!ht]
    \centering
    \includegraphics[height=6.5cm]{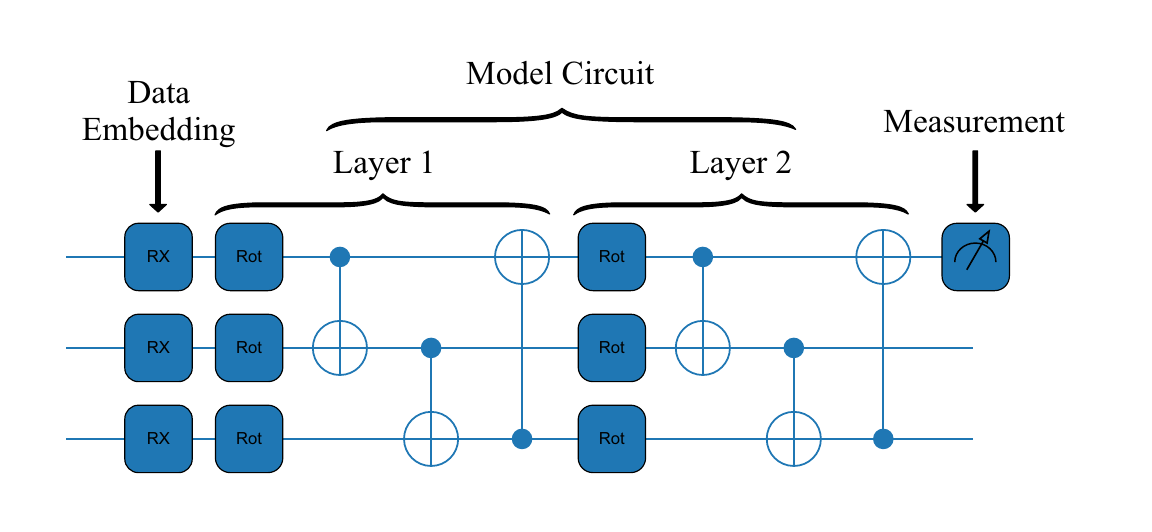}
    \caption{An example circuit for the VQC architecture used. It is comprised of 2 layers and 3 features as input. The three main stages of a QML model can be seen: embedding of the data, passing the data through the model circuit, and the measurement of the outcome.}
    \label{fig_diferent_circuits}
\end{figure*}

The output of the model is obtained by measuring the expectation value of the Pauli $\hat{\sigma}_z$ operator in one of the qubits of the final state $\psi_X'$. An example of the implementations of a VQC is represented in \autoref{fig_diferent_circuits}.

\section{Classical Machine Learning Methods}

Shallow CML methods are used to provide a baseline comparison to the QML models. The specific methods chosen for the comparison are Logistic Regression (LR) and Support Vector Machines (SVM), with these algorithms being trained with the same data as the QML algorithms.

All the classical methods were implemented using \textit{scikit-learn}~\cite{scikit-learn} library and, if not specified otherwise, the default parameters were used.

\subsection{Logistic Regression}

Logistic Regression is one of the simplest ML models and can be formulated as one of the basic building blocks of a neural network - a single-layer perceptron. The goal is to find the best set of weights $w$ that fit the data $x$:
\begin{equation}
    \hat{y}(w, b, x) = \sigma( w \cdot x + b) \, ,
\end{equation}
where $\hat y$ is the probability of an event to belong to class $1$, $w$ and $b$ are learnable parameters, and $\sigma$ is the sigmoid function.

The learning process is guided by minimizing the loss function, which in our case is the binary cross-entropy:
\begin{equation}
    L = - \mathbb{E}_x[ y \log(\hat y) + (1-y) \log(1-\hat y)] \, ,
\end{equation}
where $y$ is the binary label of whether the event is of the class signal or not, and $\mathbb{E}_x$ is the expectation value over the training data, obtained using the event weights corresponding to each signal and background process.

\subsection{Support Vector Machine}
An SVM classifier is trained by finding the hyperplane that best separates two classes of data in the hyperspace of features. It does so by using support vectors, which are the data points from the two classes closer to the hyperplane, influencing the position and orientation of the hyperplane.

The loss function of an SVM revolves around the goal of maximizing the margin, \emph{i.e} the distance between the hyperplane and the nearest data point from either class. In other words, the goal is to find the hyperplane with the greatest possible margin between itself and any point within the training set, giving a greater chance of new data being classified correctly.

Just like the Logistic Regression, the base SVM classifier can only learn a linear decision boundary. However, classification problems are rarely simple enough for it to be separable using a hyperplane, thus usually requiring a non-linear separation. SVM can do this by transforming the data using a non-linear function, named kernel, after which it can be split by a hyperplane. For this implementation, the radial-basis function (RBF) was used as kernel. This endows the SVM with a non-linear mapping where it better separates the two classes using a hyperplane.

\section{Dataset}
\label{sec_dataset}

The dataset used in this work\cite{crispim_romao_miguel_2021_5126747} is comprised of simulated events of $pp$ collisions at 13~TeV, in final states with 2 leptons, at least 1 $b$-jet, at least 1 large-$R$ jet and large scalar sum of transverse\footnote{The transverse plane is defined with respect to the proton colliding beams.} momentum
($p_T$) of all reconstructed particles in the event ($H_T$ > 500 GeV). Such basic selection corresponds to a topology commonly used in different searches for BSM events at the LHC~\cite{CrispimRomao:2020ucc}. The dominant SM background for this topology, $Zb\bar b$, and the BSM signal corresponding to $t\bar t$ production with one of the top-quarks decaying via a flavour changing neutral current decay $t\to qZ$ ($q=c,u$)~\cite{Durieux:2014xla}, were considered. Such signal was chosen given the kinematic similitude to the background, thus providing a good benchmark for the present study.

Both samples were generated with \textsc{MadGraph5 2.6.5}~\cite{madgraph} and \textsc{Pythia 8.2}~\cite{pythia}, and the detector was simulated using \textsc{Delphes 3}~\cite{delphes} with the default CMS card. Jets were clustered using the anti-$k_t$ algorithm~\cite{aktjets}, implemented via \textsc{FastJet} \cite{Cacciari:2011ma}, with $R$-parameters of 0.5 and 0.8 (the latter for the large-$R$ jets).

The following features were used for training of both the classical and quantum machine learning algorithms:
\begin{itemize}
    \item $(\eta, \phi, p_T, m, b$-tag)  of the 5 leading jets, ordered by decreasing $p_T$, with $b$-tag being a Boolean variable indicating if the jet is identified as originating from a $b$-quark by the $b$-tagging algorithm emulated by \textsc{Delphes};
    \item $(\eta, \phi, p_T, m)$  of the leading large-$R$ jet;
    \item $N$-subjettiness of leading large-$R$ jet, $\tau_n$ with $n=1,...,5$~\cite{thaler2011identifying}.
    \item $(\eta, \phi, p_T)$ of the 2 leading leptons (electrons or muons);
    \item transverse momentum (\MET) and $\phi$ of the missing transverse energy;
    \item multiplicity of jets, large-$R$ jets, electrons and muons;
    \item $H_T$.
\end{itemize}

The proportion of signal and background events was kept the same as the original simulated data during training, being $13\%$ and $87\%$ respectively. Additionally, the Monte Carlo weights, corresponding to the theoretical prediction for each process at target luminosity of 150~fb$^{-1}$, were taken into account in the evaluation of all the considered metrics and loss functions.

\section{Feature Selection}

As described in the previous section, a total of 47 features are available for training.  Considering the type of data embedding chosen, 47 qubits would be needed to train a VQC using all the dataset features. Such number of qubits is impractical given the currently available quantum computers and thus it is not feasible to train a VQC using all the features in our dataset. For the purposes of the current study, quantum computers with only 5 qubits were considered and two methods for feature selection were implemented: principal component analysis (PCA) and sequential feature selection (SFS). 

\begin{figure}[H]
    \centering
     \makebox[\textwidth]{\includegraphics[width=7cm]{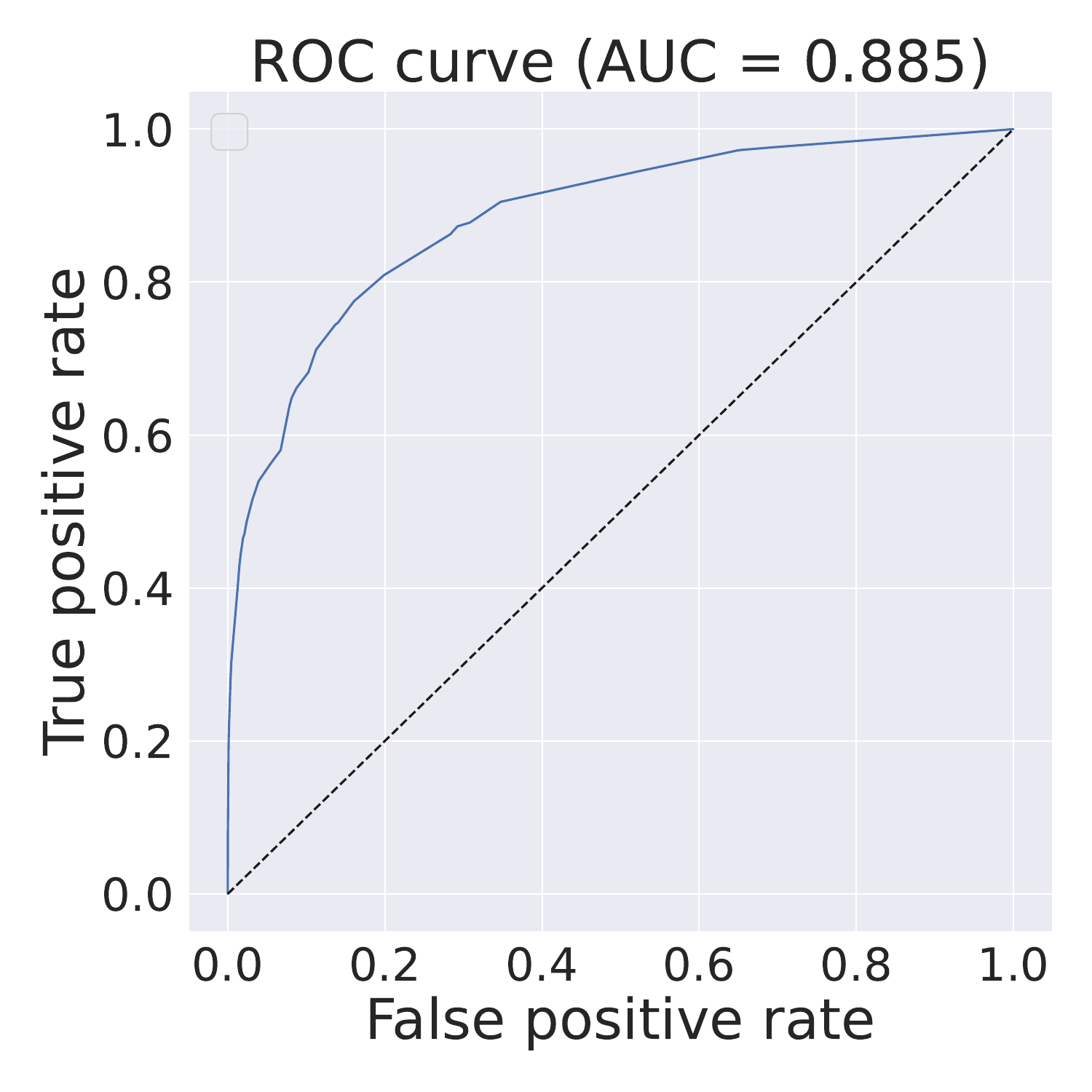}} 
    \caption{Obtained ROC curve and respective AUC score on the test dataset when training an Boosted Decision Tree, implemented with~\emph{xgboost}~\cite{Chen:2016:XST:2939672.2939785} using the full set of features and data points. The classifier has an identical configuration as the one described in \autoref{subsec:SFS}.}
    \label{fig_xgb_baseline}
\end{figure}

A relative comparison of the best 5 features~\footnote{The area under the curve (AUC) of the receiver operating characteristic curve (ROC) is considered as metric for these comparisons.} is shown in \autoref{tab:top5_original_features} while the best performance obtained with state-of-the-art CML methods without any features or data points restrictions can be seen in \autoref{fig_xgb_baseline}.

\begin{table}[H]
\centering
\begin{tabular}{|c|c|}
\hline
\rowcolor[HTML]{C0C0C0} 
{\color[HTML]{000000} Feature}        & {\color[HTML]{000000} AUC}      \\ \hline
\rowcolor[HTML]{FFFFFF} 
{\color[HTML]{000000} \MET} & {\color[HTML]{000000} 0.817} \\ \hline
\rowcolor[HTML]{FFFFFF} 
{\color[HTML]{000000}  Lepton$_1$ $p_T$}    & {\color[HTML]{000000} 0.692} \\ \hline
\rowcolor[HTML]{FFFFFF} 
{\color[HTML]{000000} Lepton$_2$ $p_T$}    & {\color[HTML]{000000} 0.649} \\ \hline
\rowcolor[HTML]{FFFFFF} 
{\color[HTML]{000000} large-$R$ jet $m$}  & {\color[HTML]{000000} 0.609} \\ \hline
\rowcolor[HTML]{FFFFFF} 
{\color[HTML]{000000} large-$R$ jet $\tau_1$}  & {\color[HTML]{000000} 0.576} \\ \hline
\end{tabular}
\caption{Top 5 features ranked by their AUC Score on the training dataset.}
\label{tab:top5_original_features}
\end{table}

\subsection{Sequential Feature Selection}
\label{subsec:SFS}

SFS algorithms are a widely used family of greedy search algorithms used for automatically selecting a subset of features that is most relevant to the problem. This is achievable by removing or adding one feature at a time based on the classifier performance until a feature subset of the desired size, $k$, is reached.

There are different variations of SFS algorithms but for the current paper, the Sequential Backward Selection (SBS) algorithm was chosen. This algorithm starts with the full set of features ($n=47$) and, at each iteration, it generates all possible feature subsets of size $n-1$ and trains a ML model for each one of the subsets. The performance is subsequently  evaluated and the feature that is absent from the subset of features with the highest performance metric is removed. This process is iterated until the feature subset contains $k$ features.

This technique was used to find subsets of 1 to 5 features. The ML model assisting the SBS was a boosted decision tree (BDT) with a maximum number of estimators set at $100$ and a learning rate of $1\times 10^{-5}$. The considered loss function was a logistic regression for binary classification and the AUC score was used as evaluation metric. The BDT was implemented using \textit{xgboost}~\cite{Chen:2016:XST:2939672.2939785} and the SBS algorithm using \textit{mlxtend}~\cite{raschkas_2018_mlxtend}. The selected features for the different values of $k$ is shown in \autoref{table_SFS} and the AUC scores for each feature in \autoref{tab:SBS_AUC}. It should be noted that \autoref{table_SFS} shows the features selected with the SBS algorithm and \autoref{tab:SBS_AUC} shows the AUC value of each one of these features. The latter is ordered by descending AUC value.

\begin{table}[H]
\centering
\begin{tabular}{|c|l|}
\hline
\rowcolor[HTML]{C0C0C0} 
$k$ & \multicolumn{1}{c|}{\cellcolor[HTML]{C0C0C0}Selected Features} \\ \hline
1 & \MET                                              \\ \hline
2 & \MET, Number of muons                                    \\ \hline
3 & \MET, Number of muons, Jet$_1$ $b$-tag                        \\ \hline
4 & \MET, Number of muons, Jet$_1$ $b$-tag, Jet$_2$ $p_T$            \\ \hline
5 & \MET, Number of muons, Jet$_1$ $b$-tag, Jet$_2$ $p_T$, large-$R$ $\tau_3$  \\ \hline
\end{tabular}
\caption{List of the features selected by the SBS algorithm for $k=1, ..., 5$.}
\label{table_SFS}
\end{table}

\begin{table}[H]
\centering
\begin{tabular}{|c|c|}
\hline
\rowcolor[HTML]{C0C0C0} 
{\color[HTML]{000000} Feature}        & {\color[HTML]{000000} AUC}      \\ \hline
\rowcolor[HTML]{FFFFFF} 
{\color[HTML]{000000} \MET} & {\color[HTML]{000000} 0.817} \\ \hline
\rowcolor[HTML]{FFFFFF} 
{\color[HTML]{000000} Number of muons}    & {\color[HTML]{000000} 0.534} \\ \hline
\rowcolor[HTML]{FFFFFF} 
{\color[HTML]{000000} Jet$_1$ $b$-tag}     & {\color[HTML]{000000} 0.418} \\ \hline
\rowcolor[HTML]{FFFFFF} 
{\color[HTML]{000000} large-R jet $\tau_3$}  & {\color[HTML]{000000} 0.316} \\ \hline
\rowcolor[HTML]{FFFFFF} 
{\color[HTML]{000000} Jet$_2$ $p_T$}       & {\color[HTML]{000000} 0.313} \\ \hline
\end{tabular}
\caption{Features selected by the SBS Algorithm and their respective AUC Score on the training dataset.}
\label{tab:SBS_AUC}
\end{table}

\subsection{Principal Component Analysis}

The PCA transforms a highly correlated, high-dimensional dataset and into a new one with reduced dimensionality and uncorrelated features, by rotating the dataset in the direction of the eigenvectors of the dataset covariance matrix. In the present paper, the PCA was performed only to remove the correlation between the features, maintaining the same dimensionality as the original data. The PCA transformation was learned from the training dataset and then applied to all datasets. When training a VQC for a specific number of features, the PCA components were ranked by AUC score and thus selected from the highest to the lowest. This is done by introducing a priority queue, \emph{i.e.} if training a model using two features is desired, the 2 top-ranked PCA components will be selected. The \textit{scikit-learn} PCA implementation was used and the obtained 5 better components are shown in \autoref{tab:pca_auc}.

\begin{table}[H]
\centering
\begin{tabular}{|c|c|}
\hline
\rowcolor[HTML]{C0C0C0} 
{\color[HTML]{000000} PCA Component} & {\color[HTML]{000000} AUC}      \\ \hline
\rowcolor[HTML]{FFFFFF} 
{\color[HTML]{000000} Component 3}   & {\color[HTML]{000000} 0.726} \\ \hline
\rowcolor[HTML]{FFFFFF} 
{\color[HTML]{000000} Component 14}  & {\color[HTML]{000000} 0.606} \\ \hline
\rowcolor[HTML]{FFFFFF} 
{\color[HTML]{000000} Component 5}   & {\color[HTML]{000000} 0.565} \\ \hline
\rowcolor[HTML]{FFFFFF} 
{\color[HTML]{000000} Component 41}  & {\color[HTML]{000000} 0.563} \\ \hline
\rowcolor[HTML]{FFFFFF} 
{\color[HTML]{000000} Component 43}  & {\color[HTML]{000000} 0.560} \\ \hline
\end{tabular}
\caption{Top 5 PCA components obtained with the training dataset, ranked by their AUC.}
\label{tab:pca_auc}
\end{table}

\section{Dataset Size Reduction}
\label{sec:dataset}

The present paper addresses the use of reduced datasets to overcome the limitation of NISQ processors while minimizing the loss of information and thus avoiding a performance loss of the QML algorithms in the HEP context. The primary method used for this purpose in the current study is \textit{KMeans}, where the $k^{th}$ most representative points, \emph{i.e.} a set of \textit{centroids}, is obtained from the original dataset. Although these \textit{centroids} are the most representative data points, they are not necessarily contained in the original dataset and, consequently, a resampling process, allowing to choose points of the original dataset (centrus), is required. 

A study of the performance of the proposed dataset reduction method will be done by training a logistic regression model with the original dataset and comparing the results with those obtained when \textit{Kmeans} and randomly undersampled datasets are used.

\subsection{KMeans Algorithm}
\label{KMeans}

Considering a clustering algorithm, \textit{Kmeans} iteratively tries to separate data into independent groups \cite{kmeans}. This separation is done using the \textit{Lloyd’s algorithm}~\cite{lloyd}, based on the minimal variability of samples within each cluster. The \textit{KMeans} algorithm requires the specification of the desired number of clusters ($k$) a priori. The following steps were used:
\begin{enumerate}
    \item \textbf{Initialization of the centroids:} using the \textit{scikit-learn} implementation, it is possible to do it in two different ways, \emph{random} and \emph{k-means++} \cite{centroid_ini}:
    
    \begin{itemize}
    \item \textbf{Random:} $k$ random samples of the original dataset are chosen.
    
    \item \textbf{K-means++:} $k$ samples of the original dataset are chosen based on a probabilistic approach, leading to the centroids being initialized far away from each other.
    \end{itemize}
    
    Assuming there is enough time, the algorithm will always converge, although the convergence to an absolute minimum is not guaranteed. The \textit{K-means++} initialization helps to address this issue. 
    Furthermore, for both initializations, the algorithm, by default, runs several times with different centroid seeds, with the best result being the output.    
    
    \item \textbf{Assignment:} Each data point $x_{i:}$ is addressed to a cluster $c_{k'}$, in such a way that the \textit{inertia} is minimized:
    \begin{equation}
        k'= \mathop{argmin}_{k}\Biggl\{\sum_{j=0}^{F-1} (x_{ij}-\mu_{kj})^2 \Biggr\},
        \label{assig}
    \end{equation}
    where $F$ is the dimensionality, \emph{i.e.} the number of features, $\mu_k$ is the centroid of the cluster $c_k$ and $j$ stands for the $(j+1)^{th}$ feature.
  
    \item \textbf{Update of the centroids' position:} The new centroids are just the means positions of each cluster, \emph{i.e.}
    \begin{equation}
        \mu_{k:}= \frac{\sum_{i=0, x_{i:}\in c_k}^{n_k-1} x_{i:}}{n_k},
        \label{mu}
    \end{equation}
    with $n_k$ being the number of samples addressed to $c_k$. It should be noted that if $n_{k'}=0$ the centroid $\mu_{k'}$ doesn't change.
    
    \item \textbf{Iteration:} Steps 2 and 3 are repeated until the maximum number of iterations is reached or until the result converges, \emph{i.e.} the centroids don't change.
    
\end{enumerate}

The \textit{KMeans} algorithm was used separately for the signal and background samples, with the corresponding weights being used.

\subsection{Dataset Resampling}
\label{Resampling}

As previously mentioned, although centroids are the most representative points, they are not necessarily contained in the original dataset. Hence, it was chosen to consider 10 neighbors of each centroid to determine each centrus, \emph{i.e.} the 10 nearest points of the original dataset.

The position of each centrus was determined using the weighted mean of the position of the neighbors,
\begin{equation}
    W = \frac{\sum_{i=0}^{9} x_{i:}\times w_i}{\sum_{i=0}^{9} w_i},
    \label{mean}
\end{equation}
where $W$ is the mean position, $x_{i:}$ is the $(i+1)^{th}$ nearest point and $w_i$ the weight of the sample.

The sample weight of each centrus was calculated based on the number of samples of the same label (\emph{i.e.} signal or background) on the original dataset:
\begin{equation}
    w_i = \frac{1}{n},
    \label{w}
\end{equation}
with $w_i$ being the weight of the $(i+1)^{th}$ centrus and $n$ the number of samples in the original dataset with the same label of this centrus.

\section{Quantum and Classical Machine Learning Training}

The training of the QML algorithms used in the current paper requires the use of optimizers. Two different ones were considered: \textit{Adam}~\cite{kingma2014adam} and tree-structured Parzen estimator sampler (\textit{TPE})~\cite{bergstra2011algorithms, bergstra2013making}.

The \textit{Adam} optimizer uses an extension of stochastic gradient descent, leveraging techniques such as adaptive moment estimation, being extensively used in optimization problems, namely in the context of machine learning. Nonetheless, since there is no reason to expect, \emph{a priori}, that it will work equally well in the context of QML, where specific challenges are expected, the \textit{TPE} optimizer was also tested.

The \textit{TPE} is a Bayesian optimization algorithm first developed for HP tuning in the context of machine learning. In the current study, it will be used to optimize VQC weights in a way very similar to what is typically done for HP tuning. \textit{TPE} is implemented using~\textit{Optuna}~\cite{optuna2019}, a library focused on HP optimization for machine learning models. \textit{TPE} works by choosing a parameter candidate that maximizes the likelihood ratio between a Gaussian Mixture Model (GMM) fitted to the set of parameters associated with the best objective values, with another GMM being fitted to the remaining parameter values. In the context of HEP, \textit{TPE} has also been used to explore parameter spaces of BSM models~\cite{deSouza:2022uhk}.

Different machine learning methods were optimized, namely a LR, a SVM and a VQC. The corresponding training was done for the set of HP summarized in \autoref{tab_HP}, where the scanned values are also listed. For each set of HP, 5 models were trained on 5 different subsets of the initial dataset (random sampling). 

\begin{table}[H]
\centering
\begin{tabular}{
>{\columncolor[HTML]{FFFFFF}}l 
>{\columncolor[HTML]{FFFFFF}}l }
\hline
{\color[HTML]{000000} Variable HP} & {\color[HTML]{000000} Possible values} \\ \hline
{\color[HTML]{000000} Feature Selection}    & {\color[HTML]{000000} {[}PCA, SBS{]}}         \\
{\color[HTML]{000000} Number of Data points} & {\color[HTML]{000000} {[}100, 500, 1k, 5k{]}} \\
{\color[HTML]{000000} Number of Features}   & {\color[HTML]{000000} {[}1, 2, 3, 4, 5{]}}        \\
{\color[HTML]{000000} Number of Layers}    & {\color[HTML]{000000} {[}1, 2, 3, 4, 5{]}} \\ \hline\hline
{\color[HTML]{000000} Fixed HP}    & {\color[HTML]{000000} Fixed values} \\ \hline
Max Epochs                         & 500                                    \\
Batch Size                         & Size of the dataset                                    \\
Learning Rate (LR)                      & 0.03                                   \\ \hline
\end{tabular}
\caption{List of scanned hyperparameters. The LR parameter is used only for the VQC optimized by Adam while the number of layers is only used by the VQCs.}
\label{tab_HP}
\end{table}

For both optimizers, the considered cost function used is the squared error, with the individual Monte Carlo samples being properly weighted.
During the training of VQCs, the inference was done on the validation dataset at 5 epoch intervals, the AUC computed and only the best-performing model, according to the previously mentioned metric, was considered.

\subsection{\textit{Adam} Implementation Details}

The training starts with the initialization of the weight vector. This is done randomly with an order of magnitude of $10^{-2}$, which is followed by training iterations until a maximum number of epochs is reached. At each iteration, the model is inferred with the training dataset, the cost function calculated and the model parameters updated via the \textit{Adam} optimizer. A summary of \textit{Adam}-optimized VQC training is shown in \autoref{algo_adam}.

\vspace{5pt}
\begin{algorithm}
\label{alg_adam}
\caption{\textit{Adam} Training}
\begin{algorithmic}
\State \texttt{$params \gets params\_initialization()$}
\For{\texttt{$epoch \gets 1$ to $max\_epochs$ }}
    \State \texttt{$loss \gets cost(params)$}
    \State \texttt{$params \gets optimizer.step()$}
    \If{\texttt{$epoch\_number \% 5 == 0$ or $epoch\_number == max\_epochs$}}
        \State \texttt{$validation\_step()$}
    \EndIf
\EndFor
\end{algorithmic}
\label{algo_adam}
\end{algorithm}

\vspace{5pt}

\subsection{\textit{TPE} Implementation Details}

We use the \textit{Optuna} implementation of the \textit{TPE} sampler. Being a Bayesian optimization algorithm, \textit{TPE} works very differently to \textit{Adam}, which is a gradient descent algorithm. In \textit{TPE}, for every training iteration, each parameter is replaced by a new value acquired sampling from a Gaussian Mixture Model of good points, which is then used to compute the loss function. At each epoch, the algorithm computes new values for the model parameters.
 With the value of the loss function of the suggested parameters, \textit{TPE} will update its internal Gaussian Mixture Models of good and bad points, which will allow it to learn what are good suggestions as more parameter values are sampled. Since \textit{TPE} is a Bayesian algorithm, it does not need to compute derivatives of the loss function, as~\textit{Adam} does, which might allow for a light workload when running trainings on quantum computers.

\section{Simulation Results}

\subsection{Feature Reduction}
\label{sim_feature_selection}

\begin{figure}
    \centering
    \makebox[\textwidth]{\includegraphics[width=16cm]{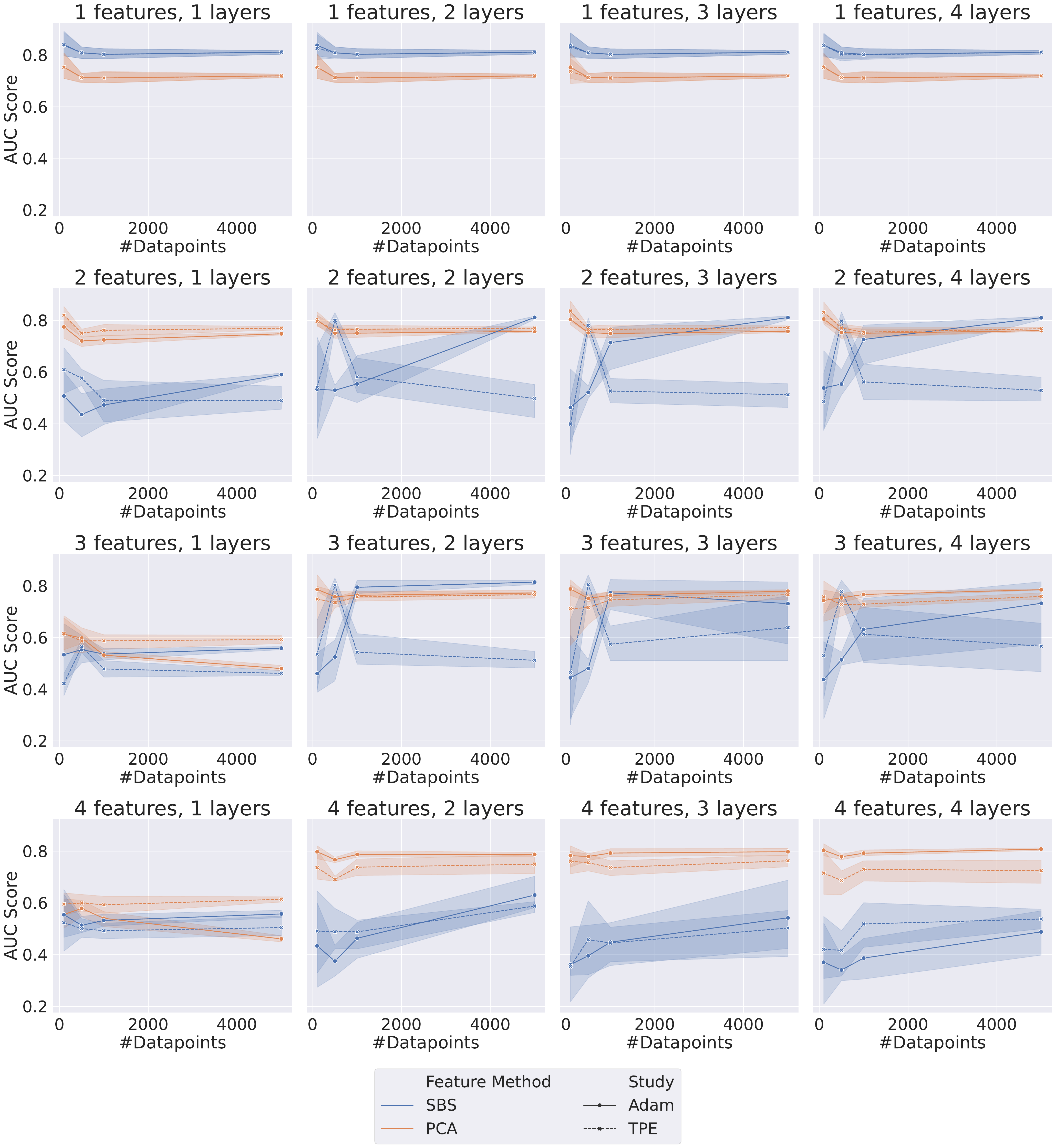}}
    \caption{Plot grid representing the results for both \textit{Adam} and \textit{TPE}-Trained VQCs. Each data point represents the AUC score on the test dataset of a different set of HP, as listed in \autoref{tab_HP}. The error bar represents the standard deviation associated with each data point since each point is the average of 5 different random samplings from the data.}
    \label{fig_adam_vs_optuna}
\end{figure}

The results indicate that QML circuits trained with SBS data are generally unstable and very susceptible to fluctuations in the randomly sampled data, as can be seen in \autoref{fig_adam_vs_optuna}.  Specifically, it is evident that using PCA-originated data produces significantly more stable results.

The performance of both optimizers, \textit{Adam} and \textit{TPE}, is usually saturated with only 2 layers. This effect is most noticeable when the number of features is greater or equal to 3. When considering only the PCA-obtained results, the two optimizers are compatible for most of the configurations tested. Exceptions occur when using a high number of features ($\geq 4$) and only one layer, where \textit{TPE} outperforms \textit{Adam}, and when using a high number of features ($\geq 4$) and more than one layer, where the opposite happens and \textit{Adam} outperforms \textit{TPE}.

\begin{figure}
    \centering
     \makebox[\textwidth]{\includegraphics[width=9cm]{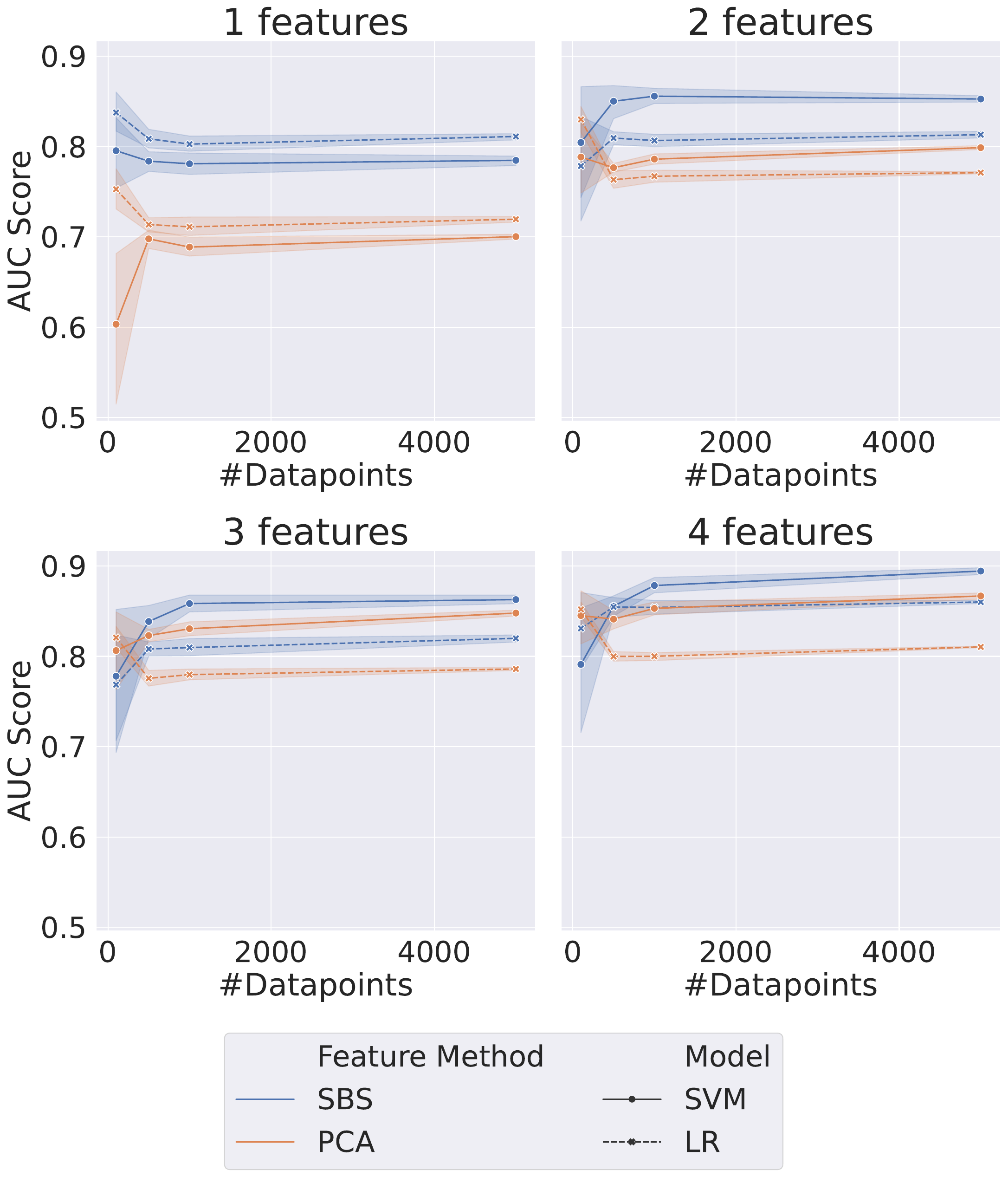}}
    \caption{Plot grid representing the results for the considered  shallow methods. Each data point represents the AUC score on the test dataset of a different set of HP, as listed in \autoref{tab_HP}. The error bar represents the standard deviation associated with each data point since each point is the average of 5 different random samplings from the data.}
    \label{fig_shallow}
\end{figure}

The shallow ML methods trained on the same data as the VQCs are shown in \autoref{fig_shallow}. The AUC scores obtained in this case are more stable for both the PCA and SBS datasets. The performance in both cases is saturated when using 2 features and the models trained with SBS data outperform the PCA-trained models, contrary to what was observed for the QML case. It should also be noted that the SVM outperforms LR in all cases except when only one feature is used, which is not surprising since SVMs are more sophisticated classifiers.

\begin{figure}
    \centering
    \makebox[\textwidth]{\includegraphics[width=10cm]{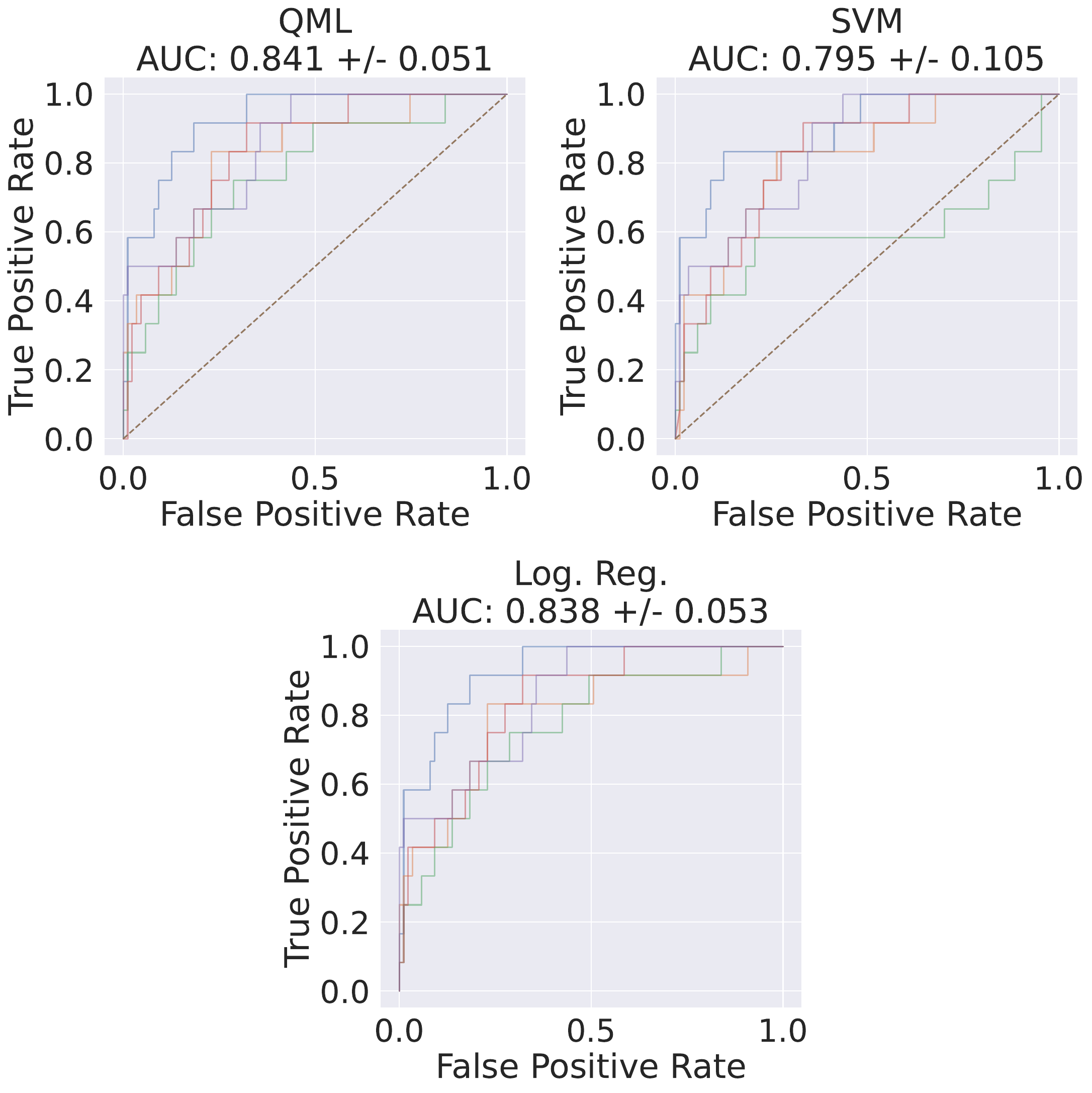}}
    \caption{ROC of the best HP set, using \textit{TPE}'s QML model average AUC score as a metric and the corresponding shallow methods ROCs for the same data. The HP for this run are SBS for feature method, 100 data points, 1 feature, and 5 VQC layers. The different colours indicate the different random samplings of the data.  }
    \label{fig_optuna_best_run}
\end{figure}

For the best set of HP, VQCs trained using \textit{TPE} and \textit{Adam} have performed similarly to the shallow ML methods (\emph{c.f.} with \autoref{fig_optuna_best_run} and \autoref{fig_adam_best_run}, respectively). It was also observed that there are no cases where QML outperforms any of the shallow methods tested. The \textit{TPE} optimizer regime produced the best performance for QML, achieving an AUC score of $0.841 \pm 0.051$, as shown in \autoref{fig_optuna_best_run}.

The reduction algorithms studied come with an additional computational cost compared to using the original dataset directly. In particular, the SBS algorithm added an overhead of 1 hour for running the XGBoost algorithm and selecting the features with more classification power. On the other hand, the PCA algorithm took a sub-minute negligible time to complete. However, since these algorithms only need to be run once, before the training, and given that the grid search for the VQC, SVM and LR algorithms took over 200 CPU hours on a dual-Intel(R) Xeon(R) Gold 6348 machine, in the end the computational cost of the classical reduction algorithms is negligible.

\subsubsection{VQC's Robustness to Discrete Features}

In the previous section it was noted that there was a significant variability in the final score of QML models, especially when training with SBS data. In fact, VQCs, being variational algorithms, are highly susceptible to small fluctuations in the data, which can have a correspondingly significant impact on the computed AUC. Additionally, numerical instabilities caused by computational floating point accuracy were observed during the validation step, leading to considerable fluctuations in the computed AUC in this regime.

\begin{figure}
    \centering
     \makebox[\textwidth]{\includegraphics[width=9cm]{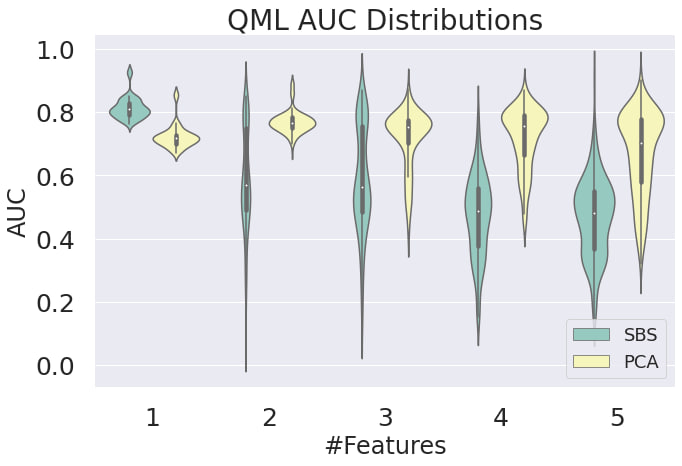}} 
    \caption{Distribution of the AUC values obtained for the QML model as a function of the number of features used in training, evaluated on the test dataset, for SBS and PCA inputs. }
    \label{violin}
\end{figure}

To further investigate this behavior, which was not observed at the same level on the PCA-trained circuits, we looked at the AUC distributions produced by QML models as a function of the number of features. As shown in \autoref{violin}, it is clear that the instability in SBS results occurs when more than two features are used.  The biggest difference in the AUC mean is found for 4 features, where the value for SBS is $0.471\pm0.129$ and for PCA is $0.719\pm0.096$. The smallest difference is found for 1 feature, where the value for SBS is $0.814\pm0.035$ and for PCA is $0.724\pm0.037$.

\begin{figure}
    \centering
    \makebox[\textwidth]{\includegraphics[width=15cm]{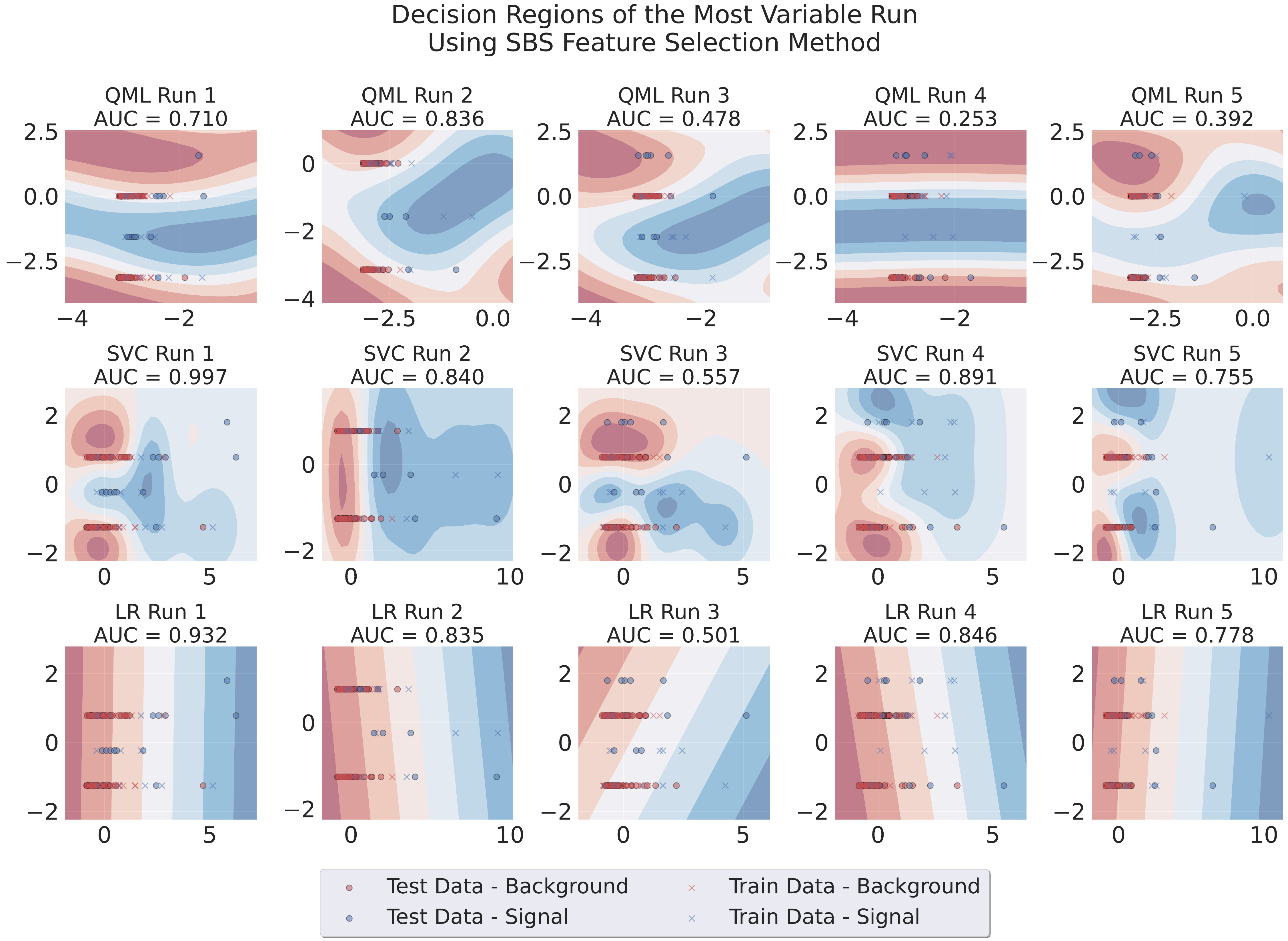}}
    \caption{Decision regions of the 3 different architectures in a run where large variability of results for the QML SBS-trained model was observed. This case uses SBS data, Adam as an optimizer, 100 data points for training and 2 layers for the circuit.}
    \label{most_variable_SBS}
\end{figure}

\begin{figure}[H]
    \centering
    \makebox[\textwidth]{\includegraphics[width=15cm]{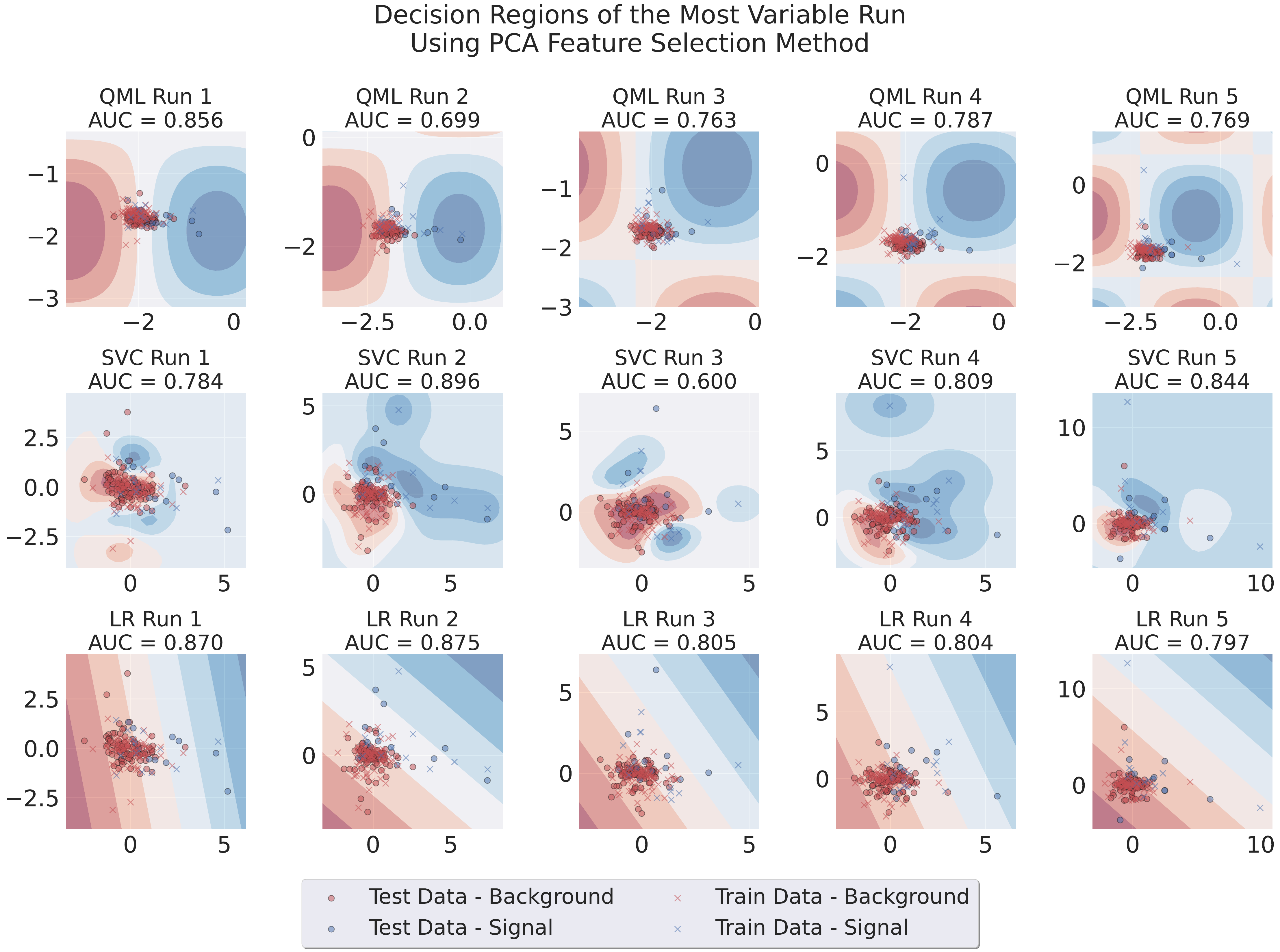}}
    \caption{Decision regions of the 3 different architectures in a run where large variability of results for the QML PCA-trained model was observed. This case uses PCA data, Adam as an optimizer, 100 data points for training and 1 layer for the circuit.}
    \label{most_variable_PCA}
\end{figure}

Additionally, we produced visualisations of the decision regions of the models trained using both feature selection methods. We focused on runs that used two features, as this is where the problem originated. \autoref{most_variable_SBS} and \autoref{most_variable_PCA} show the decision regions obtained with each model for one representative run, illustrating the sensitivity of each boundary to variations in the data, for SBS or PCA. The SBS features used are listed in \autoref{table_SFS}, where the second feature, the number of muons in the event, is a discrete variable.\footnote{In ML literature this is called a categorical variable. However, we note that even though it is categorical, it is still ordinal. As there are no non-ordinal or non-binary categorical variables in our dataset, we will refer to these variables as discrete instead of categorical for the rest of this work.}

\begin{figure}[H]
    \centering
    \makebox[\textwidth]{\includegraphics[width=16cm]{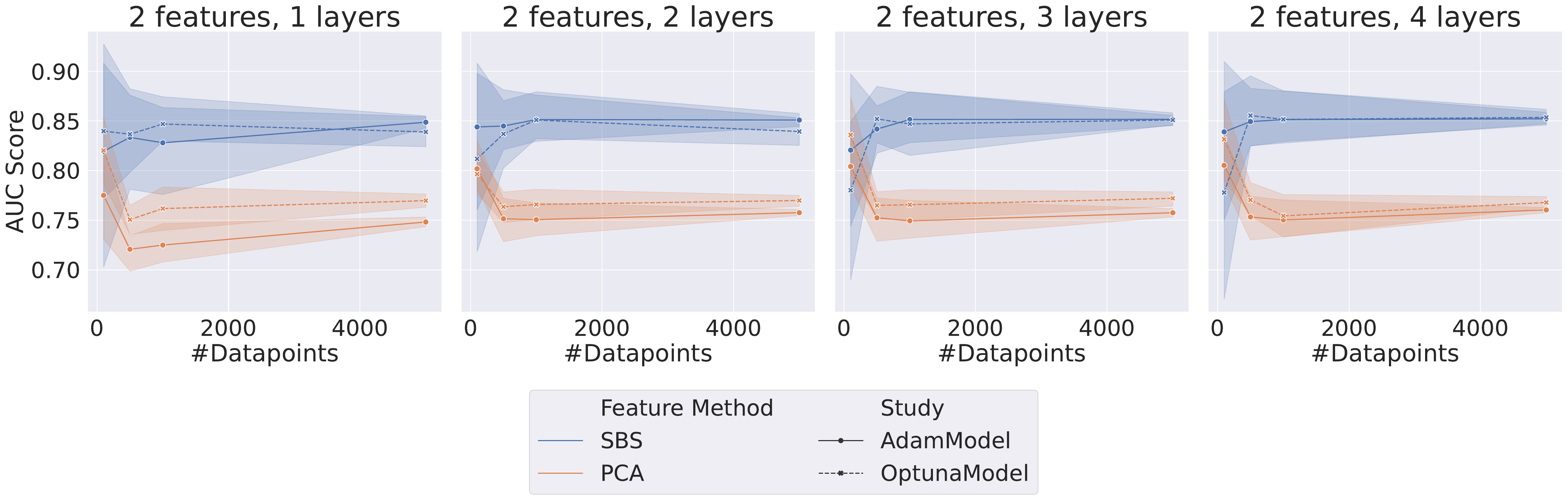}}
    \caption{Plot grid representing the results for both \textit{Adam} and \textit{TPE}-Trained VQCs. Each data point represents the AUC score on the test dataset of a different set of HP, as listed in \autoref{tab_HP}, with the two features restriction. The error bar represents the standard deviation associated with each data point since each point is the average of 5 different random samplings from the data.}
    \label{training_without_cat}
\end{figure}

While LR and SVMs are robust in the presence of discrete variables, they may pose a challenge for continuous learning algorithms such as VQCs. It is therefore possible that the variability observed when using different sub-samples of SBS data could be attributed to the use of this discrete variable. To investigate this, we conducted the SBS feature selection once again, this time excluding all discrete variables - yielding \autoref{tab:SBS_AUC_NO_CATEGORICAL}. The VQC circuits where once again trained using this modified list of inputs in a limited study of 2 features only, as illustrated in \autoref{training_without_cat}.


\begin{table}[H]
\centering
\begin{tabular}{|c|c|}
\hline
\rowcolor[HTML]{C0C0C0} 
{\color[HTML]{000000} Feature}        & {\color[HTML]{000000} AUC}      \\ \hline
\rowcolor[HTML]{FFFFFF} 
{\color[HTML]{000000} \MET} & {\color[HTML]{000000} 0.817} \\ \hline
\rowcolor[HTML]{FFFFFF} 
{\color[HTML]{000000} large-R jet $\tau_1$}  & {\color[HTML]{000000} 0.576} \\ \hline
\rowcolor[HTML]{FFFFFF} 
{\color[HTML]{000000} large-R jet $\tau_3$}  & {\color[HTML]{000000} 0.316} \\ \hline
\rowcolor[HTML]{FFFFFF} 
{\color[HTML]{000000} Jet$_2$ $p_T$}       & {\color[HTML]{000000} 0.313} \\ \hline
\rowcolor[HTML]{FFFFFF} 
{\color[HTML]{000000} Jet$_1$ $p_T$}       & {\color[HTML]{000000} 0.292} \\ \hline
\end{tabular}
\caption{Features selected by the SBS Algorithm and their respective AUC Score on the training dataset with all the discrete features removed.}
\label{tab:SBS_AUC_NO_CATEGORICAL}
\end{table}

 Using the discrete-free SBS version to train the VQC led to significantly better AUC scores, outperforming PCA-trained QML models with an average AUC score of around $0.85$, although still with larger variability than that of the PCA-trained VQCs. This is a notable departure from our previous observations in \autoref{sim_feature_selection}, where including discrete features in the SBS feature selection methodology resulted in erratic performance with no instance of outperforming PCA (except in cases where only one continuous feature was used). 
 Therefore, we found that excluding discrete variables during feature selection led to better performance for VQC circuits in a limited study of 2 features, compared to when discrete variables were included. This indicates that the choice of input features is crucial for achieving high accuracy in quantum machine learning, and future studies should consider the impact of discrete variables on VQC performance. The findings may inform future choices in selecting input features for VQC circuits to optimize model performance.

\subsection{Dataset Reduction}

\subsubsection{Implementation of \textit{KMeans}}
\label{performance}

The performance of the \textit{KMeans} algorithm was tested initially by training LR models with 10 reduced datasets and selecting a different number of $k$ features ($k\in[1, 2, 3, 4, 5]$) obtained with the SBS algorithm. The \textit{KMeans} algorithm considers the sample weight and, in order to have an equal number of signal and background centroids, it was separately applied to the signal and background data. Since state-of-the-art quantum computing requires small datasets, the data reduction studies were done for datasets with 100, 500, 1000, and 5000 data points and the number of features previously mentioned.

Two configurations were studied: the framework presented in \autoref{KMeans} was applied to the training and test datasets; and only to the training datasets (with test datasets obtained through random undersampling).\footnote{Throughout this article random undersampling refers to the random selection of data points from the original dataset. In the ML subfield of imbalanced learning, the proper methodology is to use resampling algorithms only during training, but not during validation or test. In this section we present results of these two cases as a comparison, but later we will restrict to random undersampling during validation and testing.}

\begin{figure}[H]
    \centering
     \makebox[\textwidth]{\includegraphics[width=7cm]{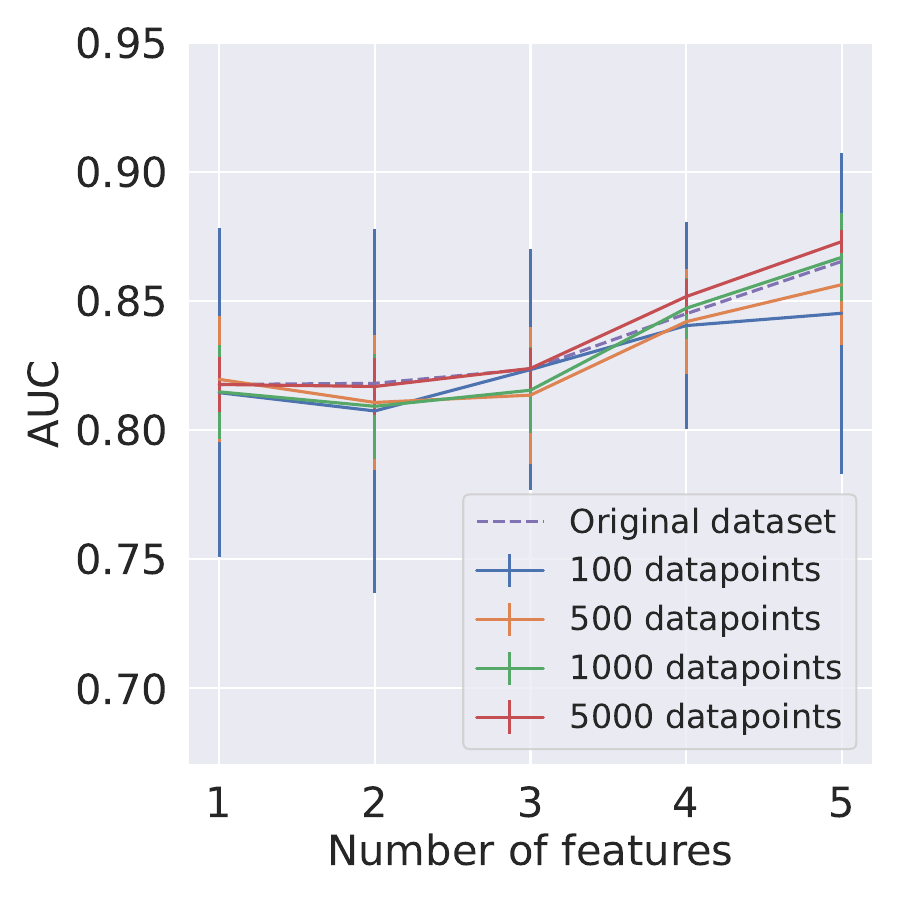}} 
    \caption{Average AUC score and corresponding standard deviation, represented as uncertainty bands, for different numbers of clusters as a function of the number of features. The training and testing datasets were reduced using the \textit{KMeans} algorithm. In each case, 10 different reduced test datasets were used.}
    \label{fig:random_undersampling}
\end{figure}

\begin{figure}[H]
    \centering
     \makebox[\textwidth]{\includegraphics[width=7cm]{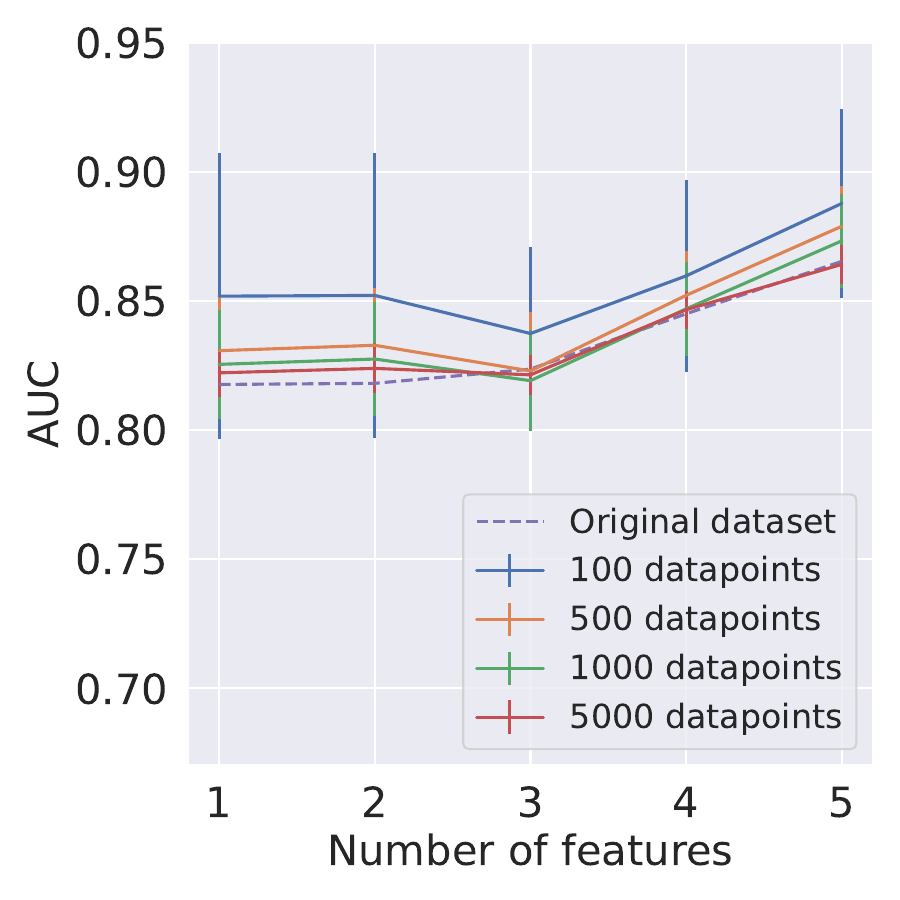}}
    \caption{Average AUC score and corresponding standard deviation, represented as uncertainty bands, for different numbers of clusters as a function of the number of features. The training dataset was reduced using the \textit{KMeans} algorithm. In each case, 10 different randomly undersampled test datasets were used.}
    \label{fig:kmeans}
\end{figure}

The mean AUC score and respective standard deviation found using \textit{KMeans} for train and test datasets are summarised in \autoref{fig:random_undersampling}. The results obtained using the \textit{KMeans} algorithm for the training dataset and random undersampling in the test signal and background samples are presented in \autoref{fig:kmeans}. In order to provide a benchmark point for comparison with the performance of the reduced datasets, a LR model was trained on the full original dataset, with results shown in both Figures.

It can be seen in \autoref{fig:random_undersampling} and \autoref{fig:kmeans} that using the \textit{KMeans} algorithm to reduce the training dataset results in AUC scores that are compatible with the performance obtained using the full original dataset.

This study shows that although \textit{KMeans} is a more sophisticated algorithm for data reduction than random undersampling, in the HEP case under consideration no significant deterioration of the performance is observed when using it, suggesting that in this study the dataset composed of prototypes is a good representative of the whole dataset in the small dataset regime, which is explored in this work.


\subsubsection{Application to QML}
\label{subsec:Perf}

The QML, SVM, and LR models were trained 
using \textit{KMeans} reduced datasets as well as random undersampling, for different dataset sizes. In this comparison, the HP for the VQC are the ones previously found to be the best, i.e. one feature chosen with the SBS method and five VQC layers for the architecture. The metric used to compare all models is the AUC score average of five different runs.

For all cases, the test and validation sets were reduced using random undersampling, hence, for each dataset size there are one train, five validation and five test datasets. The choice to keep random sampling for the test dataset, rather than \textit{KMeans} reduction, is to ensure that our methodology represents the test samples as close to the original dataset as possible, ensuring that sophisticated resampling techniques do not significantly modify the data.

\begin{figure}
    \centering
     \makebox[\textwidth]{\includegraphics[width=10cm]{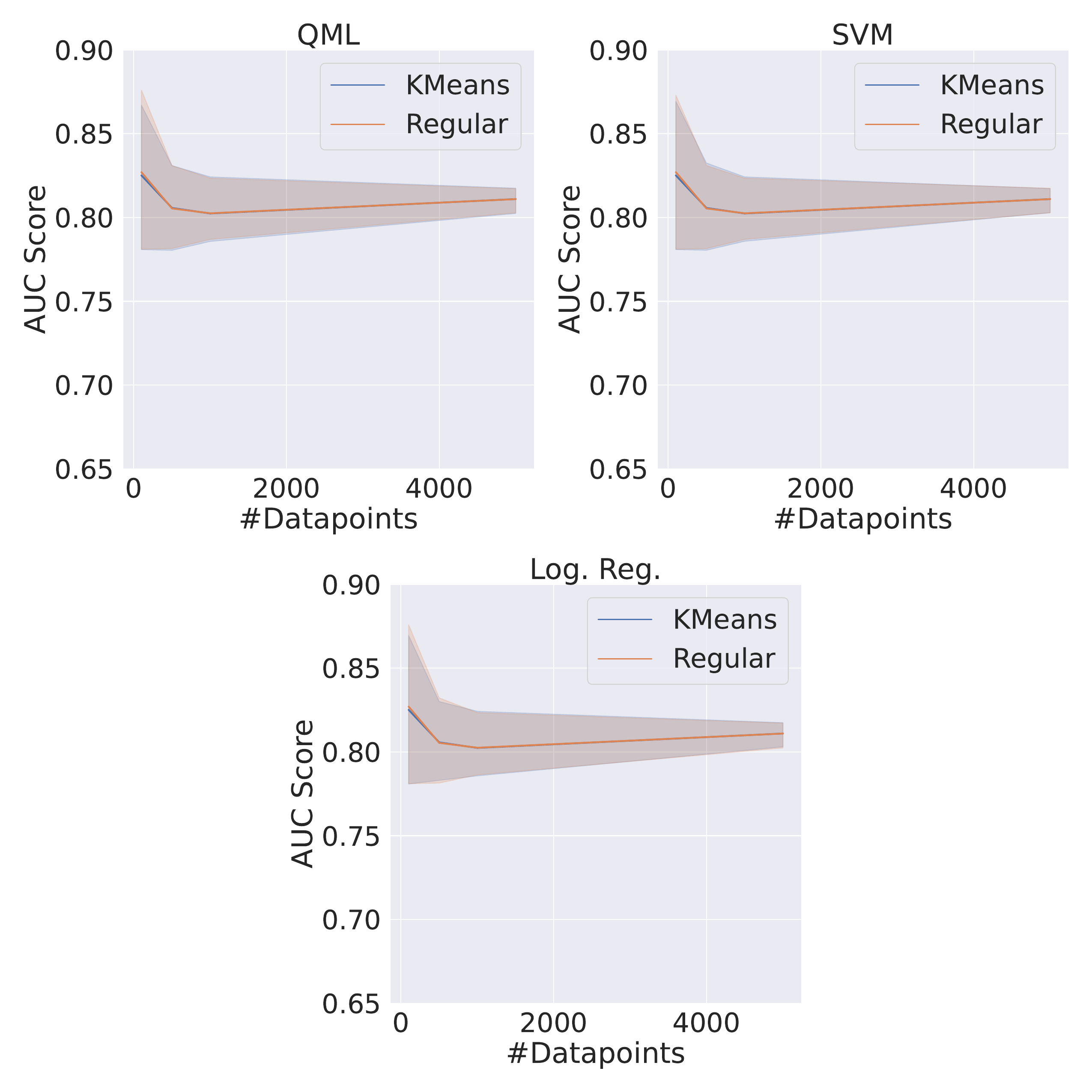}} 
    \caption{Comparison between the QML, SVM, and LR models when trained with the \textit{TPE} and the best set of HP for different dataset sizes for both random undersampling (regular) and \textit{KMeans} reduced datasets.}
    \label{fig}
\end{figure}

The obtained results are shown in ~\autoref{fig}. It can be seen that the performance for the \textit{KMeans} reduced dataset is compatible with the one obtained using the dataset reduced through random undersampling, for QML and CML models. Furthermore, the performance achieved by the simulated VQCs is identical within the statistical uncertainties to the performances by the SVM and LR, in agreement to what was observed in~\autoref{sim_feature_selection}.

Nonetheless, it should be emphasized that the model trained with random undersampling needs to be trained several times for achieving these average scores, as many times as the number of reduced datasets used. On the other hand, the models using the \textit{KMeans} reduced dataset need to be trained only once. This can be relevant in the context of quantum computers, where access is often subject to long queues and thus the number of accesses can be a limiting factor. While the \textit{KMeans} reduction technique brought an overall increase in time of around 1\%, this change is negligible taking into account the reduction in number of accesses.

\section{Real Quantum Computers Results}

\label{real_q_comp}

Until this point, only simulated quantum environments were used. In order to test the performance in real quantum computers, and thus validate the simulation results, the \textit{Pennylane} framework was used as the integration layer with \textit{Qiskit}, which works effectively as a direct API to the quantum computers provided by IBM.

\begin{figure}[H]
    \centering
     \makebox[\textwidth]{\includegraphics[width=10.0cm]{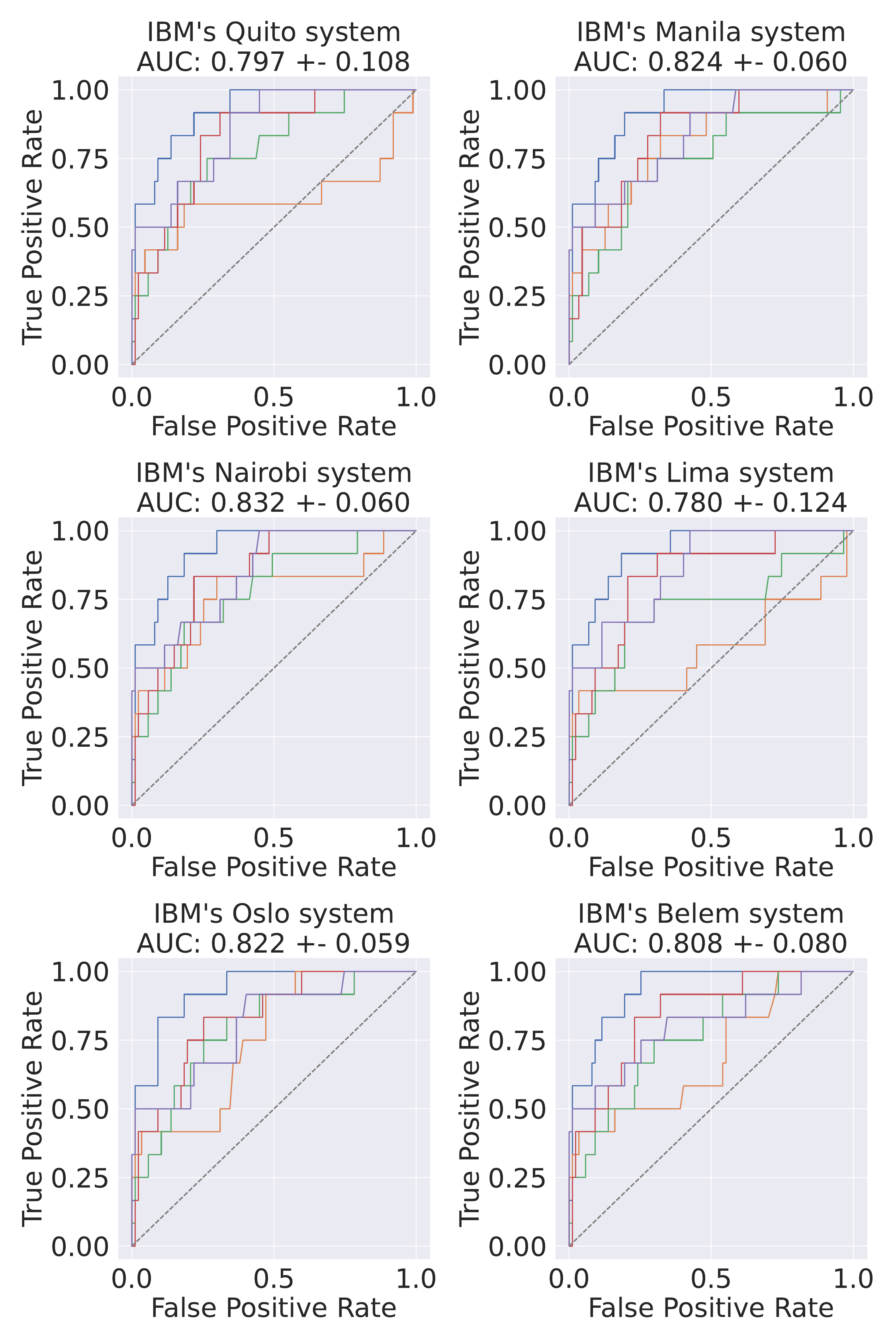}}
    \caption{Final ROC curve of the best-performing model when inferred on the test dataset in 6 different IBM systems. The average AUC scores and the corresponding standard deviations are also shown. The colours in each subplot stand for different runs of the same circuits in the same QC.}
    \label{fig_real_qc}
\end{figure}

In this study only the best performant model HP-set was used, \emph{i.e.} the \textit{TPE}-trained VQC. This VQC was implemented and its test set was inferred on six different quantum systems with identical architectures, all freely available. Evaluating our model in multiple identical quantum systems allows us to get an idea of the scale of the associated systematic uncertainty via the variability of the observed results. Since the implemented circuits are small, no error mitigation techniques were implemented. IBM's transpiler optimization level was set to 3\footnote{The level 3 of optimisation corresponds to the heaviest optimisation inherently implemented.}~\cite{Qiskit} and, for each event, the final expectation value was computed by averaging $20k$ shots on the quantum computer. The obtained results, shown in \autoref{fig_real_qc}, are compatible with the simulated ones (\autoref{fig_optuna_best_run}).

\clearpage

\section{Conclusion}

In this paper, we assessed the feasibility of using large datasets in QML applications by exploring data reduction techniques. To allow for a fair comparison between CML and QML models, we opted to 
use shallow classical 
methods as opposed to deep methods, which require large datasets that are not viable given the limitations of the current quantum computers. Our results indicate that there is comparable performance 
between CML and QML when tested on the same small dataset regime.

To achieve this, our study first compared feature selection techniques, showing that while SBS can produce the best performant QML model, it generally yielded worse and more unstable results than PCA. Additionally, we found this was produced by using discrete variables in VQCs, highlighting the suitability of PCA-transformed data for QML applications in the HEP context, where discrete variables are commonly used to describe collider events.

Our grid search over different HP combinations of VQC ran in simulation provided no evidence of quantum advantage in our study. We confirmed the results by running the best performing configuration on real-world quantum systems, obtaining compatible performances and therefore validating our conclusions. We compared the performance of TPE and Adam optimizers in QML and found that TPE achieves competitive results. Being a gradient-free optimiser, TPE offers the advantage that it can lead to faster training with a smaller memory usage when compared to Adam, which in principle can further facilitate the application of QML in current quantum computers.

We then explored data reduction techniques, finding that reducing the dataset size with the KMeans algorithm produces results that are similar to those obtained from random undersampling. This finding is significant in that it means that the model can achieve similar performance with fewer accesses to a quantum computer during training, which is a considerable bottleneck in current QML.

In conclusion, while our study found no evidence of quantum advantage in the current state of QML within the context of large HEP datasets, the performance of QML models was comparable to that of classical machine learning models when restricted to small dataset regimes. Our findings suggest that using dataset reduction techniques enables us to use large datasets more efficiently to train VQCs, facilitating the usage of current quantum computers in large datasets often found in HEP.

\hspace{10pt}

\section*{Acknowledgments}

We acknowledge the use of IBM Quantum services for this work. The views expressed are those of the authors, and do not reflect the official policy or position of IBM or the IBM Quantum team.

This work was supported by Fundação para a Ciência e a Tecnologia, Portugal, through project CERN/FIS-COM/0004/2021 
(``Exploring quantum machine learning as a tool for present and future high energy physics colliders''). IO is 
supported by the fellowship LCF/BQ/PI20/11760025 from La Caixa Foundation (ID 100010434) and by the European Union 
Horizon 2020 research and innovation program under the Marie Sk\l{}odowska-Curie grant agreement No 847648.

We thank Declan Millar, Nuno Peres and Tiago Antão for the very useful discussions and Ricardo Ribeiro for kindly providing access to some of the computing systems used in this work. We also thank Henrique Carvalho for the help in producing \autoref{fig_diferent_circuits}.

\section*{Code Availability}
The code used in this work is publicly available in \url{https://github.com/mcpeixoto/QML-HEP}.

\appendix

\section*{Appendix}

\label{a_qmlresults}

\begin{figure}[H]
    \centering
    \makebox[\textwidth]{\includegraphics[width=5cm]{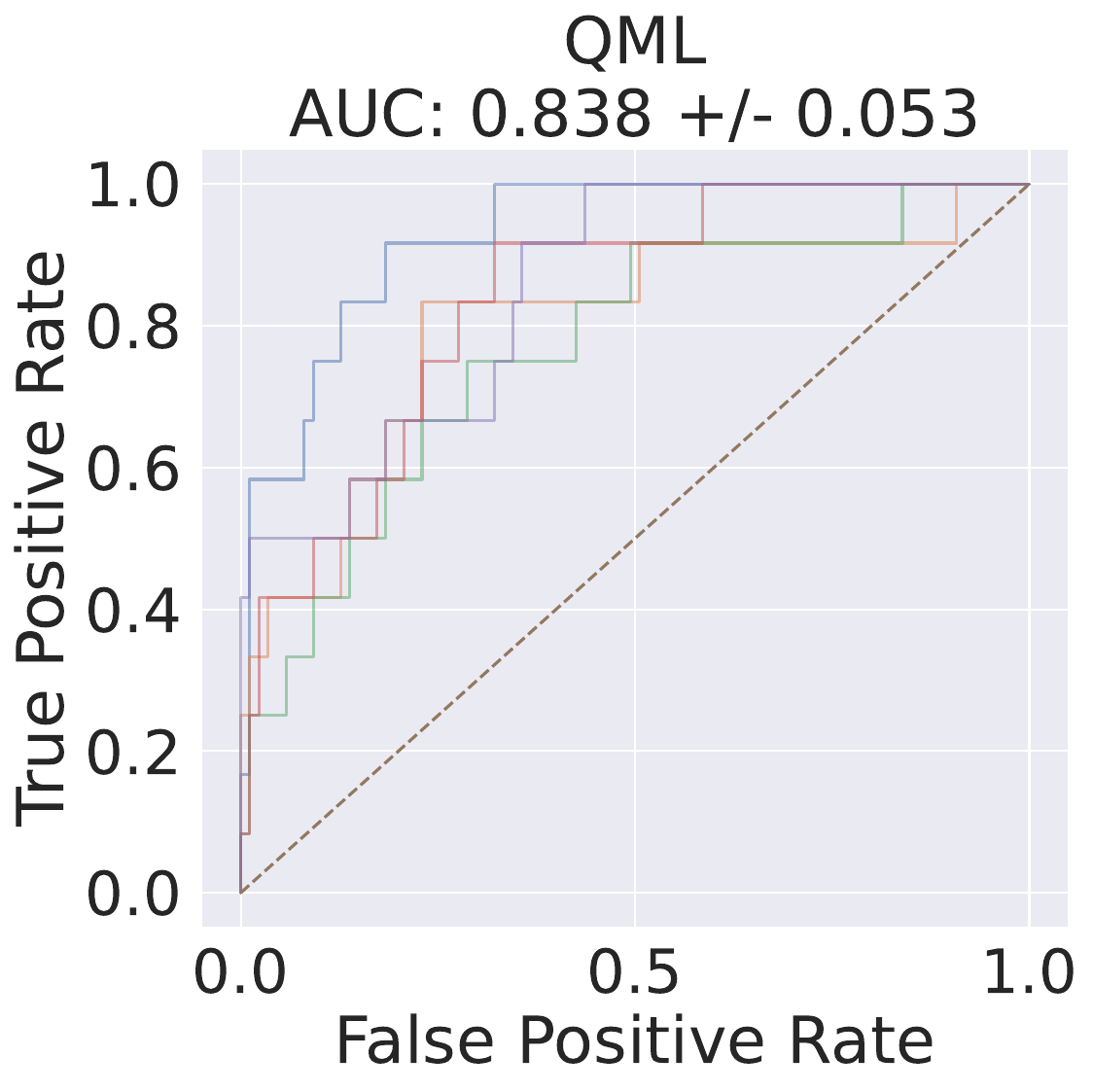}}
    \caption{ROC of the best HP set, using \textit{Adam}'s QML model average AUC score as a metric. The HP for this run are SBS for feature method, 100 data points, 1 feature, and 1 VQC layer. The corresponding shallow methods ROCs for the same data have an AUC of $0.795\pm 0.105$ for SVM and $0.838 \pm 0.053$ for LR.}
    \label{fig_adam_best_run}
\end{figure}

In this appendix we show the results obtained for the \textit{Adam}-trained VQCs, namely the ROC curve for the best HP set 
for the QML model, shown in~\autoref{fig_adam_best_run}. This result can be compared to the \textit{TPE}-trained VQC 
results shown in \autoref{fig_optuna_best_run}. As noted in~\autoref{sim_feature_selection}, the two optimizers are 
compatible.

\begin{figure}
    \centering
    \makebox[\textwidth]{\includegraphics[width=10cm]{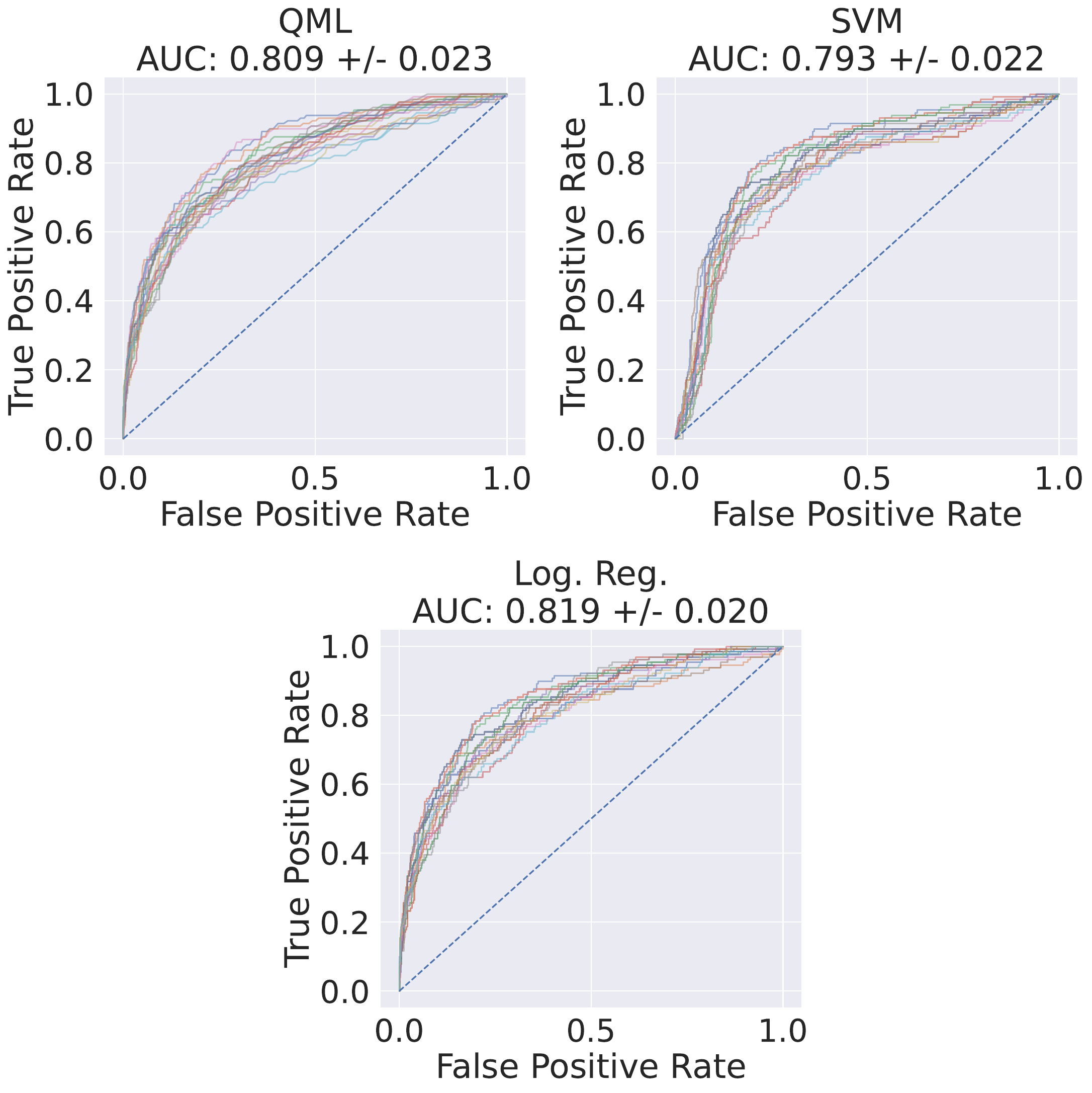}}
    \caption{ROC of the best HP set, using \textit{TPE}'s QML model average AUC score as a metric and the corresponding shallow methods ROCs. The HP for this run are SBS for feature method. 1000 data points were used. The quantum circuit was simulated considering noise. The different colours indicate each of the 20 random samplings of the data.}
    \label{fig_ROC-sim-large}
\end{figure}

In \autoref{fig_ROC-sim-large} we show the results obtained, using the simulation of a quantum circuit considering noise, with a larger number of random samplings and more data points. The \textit{Qiskit.Aer} simulator for the \textit{IBM nairobi} system was used. The corresponding noise model was assumed for the simulations and the used parameters, such as the number of qubits, number of shots, and optimization level, mirrored those considered in \autoref{real_q_comp}. The obtained results are compatible with those shown in \autoref{fig_optuna_best_run}.

\bibliography{references}{}
\bibliographystyle{SciPost_bibstyle}

\end{document}